# Coupled surface plasmon and resonant optical tunnelling in symmetric optical microcavities

Alejandro Doval, Yago Arosa, and Raúl de la Fuente

Instituto de materiais da USC - iMATUS, Grupo de Nanomateriais, Fotónica e Materia Branda, Universidade de Santiago de Compostela, E-15782, Santiago de Compostela, Spain.

**Abstract.** A symmetrical structure consisting of a low refractive index dielectric layer between two metallic films, i.e. an optical cavity, surrounded by a semi-infinite dielectric medium of higher refractive index, forms an optical system capable of supporting both volume and surface resonances. The latter are associated with synchronized collective electronic oscillations in the inner surfaces of the two thin metallic films, called coupled surface plasmons. These oscillations are generated by an evanescent wave in the cavity and therefore the thickness of the cavity is limited to the micron range for visible radiation. Under suitable incident conditions, light propagating in the microcavity will resonate with these plasmonic oscillations and can be strongly transmitted into the surrounding medium. In this work, we establish a simple model of the transmission characteristics of the cavity and define resonance conditions that allow high transmittance even for inner dielectric layer thicknesses of various wavelengths. This phenomenon is an enhanced version of the optical process of frustrated total reflection between dielectrics analogous to quantum tunnelling effect. In the present situation, the phenomenon is more striking because it takes place in a system with two absorbing metal films which, under resonant conditions, favour a superior transmission.

## 1. Introduction

The possibility of confining light to the surface between a metal and a dielectric material is well known and has been studied since the second half of the 20th century [1, 2]. If two materials in contact have real parts of their dielectric constants with opposite signs, the interaction of light with the free charges in the medium with the negative dielectric constant can cause a collective oscillation when a certain synchronism condition is fulfilled. This is the case with noble metals in contact with a dielectric material in the visible range, showing a predominantly real and negative dielectric constant, and with their conduction electrons free to couple to electromagnetic radiation. These oscillations are known as surface plasmon resonance (SPR). Taking this a step further, if a very thin metal or dielectric film is surrounded by two semi-infinite dielectric or metallic materials, respectively (DMD or MDM structure), it is possible to give rise to collective and synchronized oscillations involving the two surfaces of the thin film. These are called coupled surface plasmons (CSP).[3]

The first studies on this field focused on guided resonances or simple plasmonic modes, at a single surface or at a metallic or dielectric film. The main objective was to establish the resonance conditions and the corresponding dispersion curves [4–6]. The observation of plasmon resonances by means of prism coupling encouraged this kind of studies and led to their first applications [7]. The classical SPR coupling systems are Kretschmann and Otto configurations. Kretschmann configuration [8] consists of a high index prism (H) with a thin metallic layer (M) deposited on its hypothenuse, in contact with air or some other low index dielectric material (L), forming a HML structure. Otto configuration [9], in contrast, consists of a high index prism (H) in direct contact with a low index dielectric film (L), which, in turn, is in contact with a thick metal (M), forming a HLM structure. In both cases, the surface plasmon is produced at the ML interface, and is observed through the reflection output of the system, when attenuated total reflection (ATR) occurs.

Actually, neither the Kretschmann nor the Otto configuration involves guided resonances (modes) at the interface, but radiative resonances associated with the HML or HLM structure, respectively. This is because the resonance corresponds to a dip in reflectance, for which the missing radiation is absorbed, and not guided. Furthermore, as noted in [10], although close, reflection and SPR resonances do not coincide at the same incidence angle or wavelength. In the case of CSP radiative resonances, a good method for their observation consists in using a planar microcavity with thin metallic films acting as mirrors, with a low index intracavity medium between them, and surrounded by a high index dielectric material, i.e. a HMLMH structure with possible CSPs localized at the two ML interfaces. These plasmon resonances in microcavities were early observed through both the reflection [11] and transmission [3, 12] outputs, and consequent numerical studies were performed. Further fundamental studies of CSP resonance are scarce, and most research on the topic is focused on analysing different applications. The main applications studied are related to the design of optical filters [13–15] based on transmission resonances, or to the improvement of coupling efficiency in Kretschmann configuration [16–19] based on reflection resonances. In addition to these, there are several other applications of the structure, such as electroluminescence [20], SPR spectroscopy [21], potential hyperlenses [22, 23], thin air gap measurements [24, 25] or biosensing [26].

Here we revisit CSP resonances in optical microcavities from a fundamental perspective. We are presenting an analytical model for transmission through a plane metallic cavity, operating in the plasmonic regime (for incidence angles above the critical incidence for a HL interface). It is based on the theory for multiple beam interference [27], that we have extended to the plasmonic regime, at which the wave in the cavity is evanescent. In this case, the cavity thickness must be in the order of a micron, so that total internal reflection (TIR) does not happen at the first mirror because the corresponding tail reaches the second mirror. The metallic film must also be thin, no more than a few tens of nanometres, so that light is not totally attenuated when crossing through it. Our model allows us to establish a well-defined resonance condition for transmittance maxima, for both lossless and lossy metals, and to determine the maximum value of transmittance for different cavity thicknesses. In this study, we will find that the peculiar shape of the amplitude coefficient $r_{lmh}$ at the mirrors inside an HMLMH structure plays a key role.

## 2. Transmission through a HMLMH structure

The device we are studying consists of a set of five media separated by plane-parallel interfaces. It is a symmetric arrangement, as shown in Fig. 1, where the three central layers make up a microcavity. The structure is formed by a surrounding dielectric of refractive index $n_h$ (h stands for high), two thin metallic mirrors (which will be referred to using subscript m) and an inner dielectric of refractive index $n_l < n_h$ (l stands for low). The latter fills the space between the mirrors, which are separated by a distance $d$, referred to as the 'intracavity thickness'.

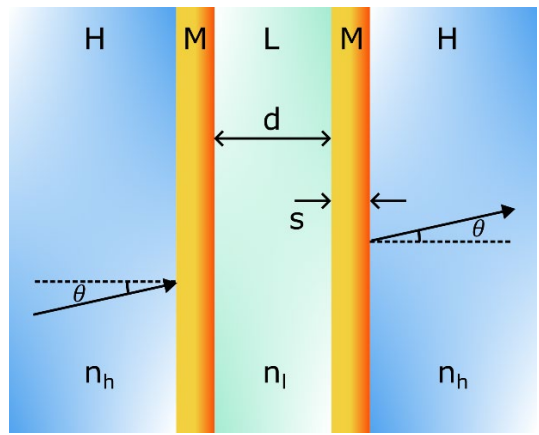

Fig. 1 Schematic drawing of an HMLMH optical structure

The index inequality, $n_h > n_l$, allows the possibility of the incidence angle exceeding a critical value $\theta_{cr} = \sin^{-1}(n_l/n_h)$. In that case of $\theta > \theta_{cr}$, there is an evanescent wave inside the cavity. This fact allows the coupling of light to different types of resonances within the structure, resulting in its transmission through the multilayer for incidence angles under which light would not have been expected to reach the other side. That means a resonant optical tunnelling process [28–30].

This atypical transmittance can be analytically modelled using Fresnel coefficients iteratively at each of the interfaces and considering the propagation of electromagnetic (EM) fields through each of the layered media [31, 32]. Besides, plasmonic resonances are associated with charge oscillations in the interface between metallic and dielectric media, requiring a non-null field component in the direction normal to these interfaces, that means, transverse magnetic (TM) or p-polarized waves. Therefore, only TM waves must be considered. Generalized Fresnel formalism leads to the following expressions for the reflection and transmission amplitude coefficients of the whole system:

$$r_{h/h} = r_{hml} + \frac{t_{hml}\, t_{lmh}\, r_{lmh}\, e^{i2k_{l\perp}d}}{1 - r_{lmh}^2\, e^{i2k_{l\perp}d}} = \frac{r_{hml} + a_{lmh}\, r_{lmh}\, e^{i2k_{l\perp}d}}{1 - r_{lmh}^2\, e^{i2k_{l\perp}d}}$$
$$t_{h/h} = \frac{t_{hml}\, t_{lmh}\, e^{ik_{ln}d}}{1 - r_{lmh}^2\, e^{i2k_{l\perp}d}} \tag{1}$$

with $h/h$ an abbreviation for $hmlmh$. Furthermore, [31, 33]:

$$a_{lmh} = a_{hml} = t_{lmh}t_{hml} - r_{lmh}r_{hml} = \frac{e^{i2k_{l\perp}d} + r_{mh}r_{ml}}{1 - r_{mh}r_{ml}e^{i2k_{l\perp}d}} \tag{2}$$

The three subindices of each coefficient in the previous formula account for the three-layered media faced by the incident EM radiation. The reflection (transmission) amplitude coefficient $r_{ijk}$ ($t_{ijk}$) corresponds to a wave arriving from medium "i" towards medium "k", with the medium "j" between those two. $k_{l\perp}$ is the normal component of the wavevector inside the cavity (within medium L), that can be calculated from the refractive indices and the incidence angle:

$$k_{l\perp} = k_0\sqrt{n_l^2 - n_h^2 \sin^2\theta} \tag{3}$$

Given the expression for the critical incidence angle, it is clear that $k_{l\perp}$ will be real for $\theta < \theta_{cr}$ and pure imaginary for $\theta > \theta_{cr}$ (only evanescent waves are allowed within the inner medium then). In any case, $r_{imk}$ and $t_{imk}$ coefficients in eq. (1) define the optical properties of the two cavity mirrors, representing amplitude reflection and transmission either from outside ($r_{hml}$ and $t_{hml}$) or from inside ($r_{lmh}$ and $t_{lmh}$) the cavity.

These can be expressed in terms of the TM Fresnel coefficients associated with the single metal-dielectric interfaces:

$$r_{imj} = \frac{-r_{mi}+r_{mj}e^{i2k_{m\perp}s}}{1-r_{mi}r_{mj}e^{i2k_{m\perp}s}} \qquad t_{imj} = \frac{t_{im}t_{mj}e^{ik_{m\perp}s}}{1-r_{mi}r_{mj}e^{i2k_{m\perp}s}} \qquad (4)$$

where $s$ represents the thickness of each metallic mirror, and $k_{m\perp}$ is again the normal component of the wavevector, this time the one inside the metal:

$$k_{m\perp} = k_0\sqrt{n_m^2 - n_h^2 \sin^2\theta} \qquad (5)$$

It should be remarked that the last terms in eqs. (3) and (5) are equal because the component of the wavevector parallel to the interfaces is an invariant under propagation in the system, $\beta = k_0 n_h \sin\theta$. The TM Fresnel coefficients for incidence at a single interface from medium "i" towards medium "j" are given by:

$$r_{ij} = \frac{\varepsilon_j k_{i\perp} - \varepsilon_i k_{j\perp}}{\varepsilon_j k_{i\perp} + \varepsilon_i k_{j\perp}} \qquad t_{ij} = \sqrt{\frac{\varepsilon_i}{\varepsilon_j}}(1+r_{ij}) \qquad (6)$$

with $\varepsilon_i = n_i^2$ the relative electrical permittivity in medium 'i'. Considering all the above, one can just square the modulus of eq. (1) to obtain the transmittance of the cavity, $T = |t_{h/h}|^2$. This formula is well known when considering small incidence angles, but the novelty here lies in its generalization for angles greater than the critical angle $\theta_{cr}$ between the two dielectric media:

$$T = \frac{|t_{lmh}t_{hml}|^2}{4\rho_{lmh}^2[\sinh^2(k_{l\perp}''d - \ln\rho_{lmh}) + \sin^2(k_{l\perp}'d + \varphi_{lmh})]} \qquad (7)$$

with $r_{lmh} = \rho_{lmh}e^{i\varphi_{lmh}}$, and where we have split the normal component of the wavevector into its real and imaginary parts, $k_{l\perp} = k_{l\perp}' + ik_{l\perp}''$. The previous formula shows a quite straightforward dependence on the intracavity thickness $d$, but more complex dependences on both the incidence angle $\theta$ and the wavelength $\lambda$. Nevertheless, extracting some relevant information is still feasible.

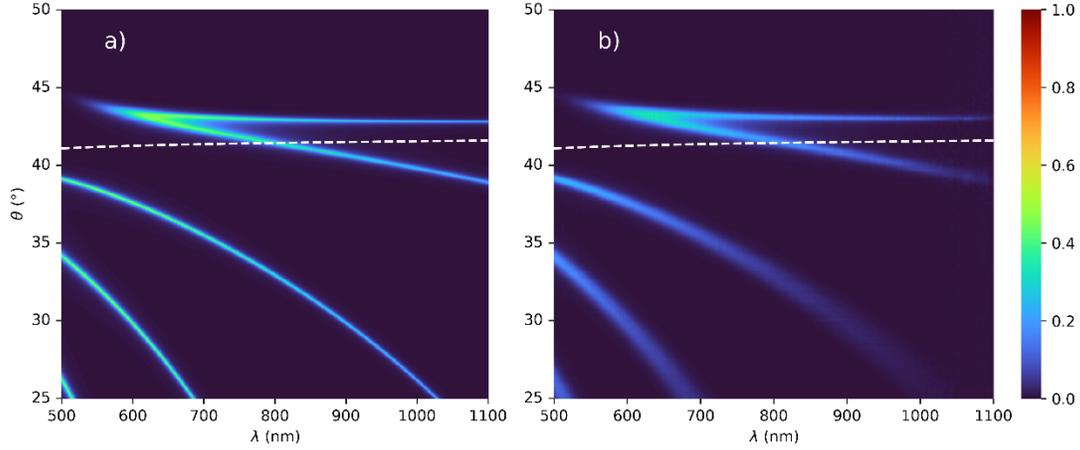

Fig. 2 Example of transmission resonance in an optical microcavity using BK7 (h), silver (m) and air (l). a) Simulation with $d = 1250\ nm$ and $s = 45\ nm$ b) Experimental. The white dotted line corresponds to the critical angle curve $\theta_{cr}(\lambda)$.

A typical transmittance map in incidence angle and wavelength for a cavity with a fixed given thickness is shown in Fig. 2. The different curves or branches that appear on the map correspond to resonances of the cavity, which are defined by the condition of maximum transmittance. This last detail is important, since the location of reflectance minima is slightly shifted, because it depends on the reflection coefficient $r_{hml}$, whereas the transmittance does not. There are two main types of resonances. On the one hand, for incidence angles below the critical angle ($\theta < \theta_{cr}$) we have photonic or volume resonances, involving harmonic waves [27]. These are Fabry-Perot type resonances for very thin cavities. On the other hand, above the critical incidence ($\theta > \theta_{cr}$) we find surface resonances, which are plasmonic in nature since they are associated with the creation of charge density oscillations at the dielectric-metal interfaces. The plasmonic resonances involve high electric fields at both ML interfaces inside the cavity, and we will therefore speak of coupled surface plasmons (CSP), as previously stated. It can be observed in Fig. 2 how they merge into a single resonance, and from that point on, the transmittance gradually vanishes. It is also clear how the second of the plasmonic resonances matches the first of the photonic resonances at the critical angle. All these resonances and their behaviours resemble the allowed modes in an MDM structure with semi-infinite metals [34, 35].

After this first example presented as an overview of the output of the microcavity, we will now start a thorough analysis of its transmission. First, we will consider a fixed wavelength, and we will study the simplest ideal case, where the permittivity of the metal is purely real.

### 3. CSP resonances in ideal lossless metals

As mentioned above, the photonic or volume resonances, allowed for incidence angles below $\theta_{cr}$, correspond to the same well-known physical phenomenon as those in a Fabry-Pérot interferometer. For this reason, our theoretical analysis will focus on the CSP resonances in the cavity, corresponding to incidence angles such that $\theta > \theta_{cr}$, the angular range that we will consider from now on. In this case, $k_{l\perp}$ is purely imaginary, and equation (T) is therefore simpler:

$$T = \frac{|t_{lmh} t_{hml}|^2}{4\rho_{lmh}^2 [\sinh^2(k_{l\perp}'' d - \ln \rho_{lmh}) + \sin^2(\varphi_{lmh})]} \qquad \theta > \theta_{cr} \qquad (8)$$

By way of illustration, we will first deal with the ideal case where the permittivity of metals follows Drude's model without losses, and therefore it is real:

$$\varepsilon_m = 1 - (\omega_p/\omega)^2 \qquad (9)$$

As has been anticipated, we will not study spectral variations at first, so the frequency has a fixed value, $\omega = \omega_0$, lower than the plasma frequency $\omega_p$. This condition, $\omega_0 < \omega_p$, means that the real permittivity of the metal is negative, i.e. $\varepsilon_m = \varepsilon_m' < 0$. Otherwise, for frequencies over the plasma value, the permittivity of both materials would be positive and no plasmons would be allowed. Given that $\varepsilon_m = n_m^2$, choosing a frequency below $\omega_p$ means a purely imaginary refractive index, i.e. $n_m = i n_m''$. Together with eq. (5), this means $k_{m\perp} = i k_{m\perp}''$ is also imaginary. Consequently, we get from eq. (6) that $\rho_{mh} = 1$, and $r_{ml}$ is real and greater than unity. This unusually high value of the Fresnel reflection coefficient between the mirrors and the inner medium leads in eq. (4) to high $r_{lmh}$ values, which appear in eq. (8) and are responsible for the resonant transmission maxima when the incident energy couples to CSPs. This will be discussed later.

Since the system is transparent, the energy flow in the cavity must match the transmitted flow. In this situation it can be shown that $|t_{lmh}|^2 = \left|\frac{k_{ln}}{k_{hn}}(r_{lmh} - r_{lmh}^*)\right|$. This equality, together with the relation $k_{h\perp} t_{lmh} = k_{l\perp} t_{hml}$ [31], leads to $|t_{lmh} t_{hml}|^2 = 4|r_{lmh}|^2 \sin^2(\varphi_{lmh})$. Therefore, the expression of transmittance in eq. (8) can be rewritten into a simplified version for incidence angles above the critical angle:

$$T = \frac{1}{1 + \left[\frac{\sinh(k_{l\perp}'' d - \ln \rho_{lmh})}{\sin(\varphi_{lmh})}\right]^2} \qquad (10)$$

According to this equation, the transmittance of the microcavity will reach a maximum value of one if the hyperbolic sine vanishes, i.e.:

$$d = \frac{1}{k_{l\perp}''} \ln \rho_{lmh} = \frac{1}{k_0 \sqrt{n_h^2 \sin^2 \theta - n_l^2}} \ln \rho_{lmh}(\theta) \equiv H(\omega_0, \theta) \equiv F(\theta) \qquad (11)$$

This condition expresses the existing constraint between the incidence angle and the cavity thickness that can only be fulfilled if $\rho_{lmh} > 1$. In general, there are two CSP resonances that are solutions of eq. (11), corresponding to two different continuous curves in the $(\theta, d)$ plane (see fig. 3a). The first, located at higher incidence angles, starts at $\theta = 90°$ for its minimum value of $d = F(90°)$ while the second begins at the critical angle (the limit of the considered angular region) for its minimum intracavity thickness $d = F(\theta_{cr})$, From those points, the two resonance curves converge to a common value of $d$, $d_{max} \equiv d_{co} = F(\theta_{co})$, at an intermediate angle $\theta = \theta_{co}$. $(\theta_{co}, d_{co})$ are what we call the coalescence angle and coalescence thickness, since they correspond to the common values at which both CSP resonance curves meet. $d_{co}$ is the maximum value of the intracavity thickness that verifies eq. (11). However, $T$ can still have maxima at $\theta = \theta_{co}$ for thicker cavities ($d > d_{co}$) despite not reaching $T = 1$, so they must be considered in our study.

By comparing with the modes of an MDM structure [35], it can be shown that the CSP at higher incidence angles corresponds to the fundamental resonance $TM_0$, that already appears for small cavity thicknesses and has a symmetric magnetic field distribution across the structure (the distribution is also symmetric for the component of the electric field normal to the interfaces). The other CSP presents an antisymmetric magnetic field distribution and corresponds to the $TM_1$ resonance.

Fig. 3a shows the curve of maximum transmittance in the $(d, \theta)$ plane for incidence angles above the critical angle. It can be seen how the two resonant solutions coalesce at $(d_{co}, \theta_{co})$ and become degenerated for greater cavity thicknesses, producing a single observable maximum in transmittance at that same incident angle $\theta_{co}$. Fig. 3b shows the maximum transmittance values for different cavity thicknesses. Full transmission takes place even through cavities with thicknesses of several wavelengths. Over the coalescence thickness, the transmittance maximum $T_{max} = T(d, \theta_{co})$ decreases gradually towards 0, because of the hyperbolic sine in eq. (10) becoming bigger. The transmittance curve can be divided into three distinct regions: i) the zone of full transmission, $d < d_{co}$, where $T_{max} = 1$; ii) the zone of FTIR with moderate transmission $d \gtrsim d_{co}$, where $T_{max} = T(d, \theta_{co}) < 1$; iii) the zone of FTIR with low exponentially decreasing transmission, $d - d_{co} \gg 1/k_{l\perp}''$, where $T_{max} = T(d, \theta_{co}) \cong \sin^2(\varphi_{lmh}) e^{-2k_{l\perp}''(d-d_{co})} \ll 1$. The limit $T_{max} = 0$ corresponds to TIR. We can compare this curve to the corresponding one for the phenomenon of FTIR in dielectric materials [36]. In this case, the transmittance is obtained from eq. (10) making the substitution $r_{lmh} \to r_{lh}$. Since $\rho_{lh} = 1$, it necessarily implies that $\ln \rho_{lh} = 0$, and there is not a region of full transmission in FTIR. However, the behaviour is similar in the two other regions.

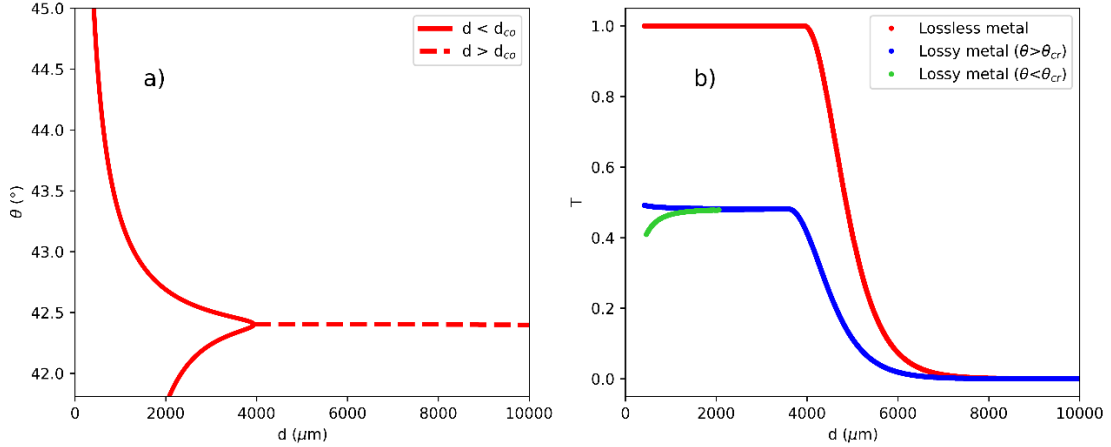

Fig. 3 In red, resonance curve and its transmittance for an ideal lossless metal with permittivity given by eq. (9) with λ = 1 μm and $\omega_p$ = 1.35 $10^{16}$ s⁻¹, $n_h = 1.5$ and $n_l = 1$. The red dotted line shows the degenerated resonance line at which both $TM_0$ and $TM_1$ resonances have coalesced for thicknesses greater than the coalescence thickness $d_{co}$. For comparison, resonant transmittance in a lossy metal (Γ = 6 $10^{13}$ s⁻¹) is also shown. The blue line in (b) corresponds to the two plasmonic resonances, while the green line corresponds to the $TM_1$ resonance in the photonic regime. Before the coalescence distance the two plasmonic resonances have similar but slightly different transmittance, which is greater for the fundamental one, and becomes degenerated at $d_{co}$.

## 4. CSP Resonances in lossy metals

Leaving the ideal case behind means considering a real metal with a complex dielectric constant, with a non-zero imaginary part, accounting for absorption. However, its negative real part is still considered to be much bigger in magnitude, i.e. $\varepsilon = \varepsilon' + i\varepsilon''$, where $-\varepsilon' \gg \varepsilon'' > 0$. This last inequality is true for most plasmonic materials in the infrared – visible spectral region and is commonly used in the modelling of plasmonic phenomena. Introducing losses in the Drude model leads to the following expression for relative permittivity, which is used in table 1 to model the permittivity for some metals:

$$\varepsilon_m = 1 - \frac{\omega_p^2}{\omega^2 + i\omega\Gamma} \qquad (12)$$

Table 1. Values of the plasma frequency and damping constant for different metals[37] and the resulting values for permittivity at visible and infrared wavelengths

| Material | $\omega_p$ ($s^{-1}$) | $\Gamma$ ($s^{-1}$) | $\varepsilon$ | |
|---|---|---|---|---|
| | | | $\lambda = 600\ nm$ | $\lambda = 1000\ nm$ |
| Silver | $1.369 \times 10^{16}$ | $2.730 \times 10^{13}$ | $-18.013 + 0.165\ i$ | $-51.807 + 0.765\ i$ |
| Gold | $1.371 \times 10^{16}$ | $4.040 \times 10^{13}$ | $-18.067 + 0.245\ i$ | $-51.948 + 1.136\ i$ |
| Copper | $1.122 \times 10^{16}$ | $1.378 \times 10^{13}$ | $-11.772 + 0.056\ i$ | $-34.476 + 0.260\ i$ |
| Aluminium | $2.240 \times 10^{16}$ | $1.242 \times 10^{14}$ | $-49.827 + 2.011\ i$ | $-139.796 + 9.283\ i$ |
| Nickel | $7.419 \times 10^{15}$ | $6.626 \times 10^{13}$ | $-4.582 + 0.118\ i$ | $-14.493 + 0.545\ i$ |

The most significant change in transmittance in comparison to the previous ideal case is that absorption at the metallic mirrors causes a reduction in the fraction of energy that is transmitted through the microcavity as shown in Fig. 3b. In this case, eq. (10) is no longer valid, but we can rewrite transmittance as follows:

$$T = \frac{T_0}{1 + \left[\frac{\sinh(k''_{l\perp} d - \ln \rho_{lmh})}{\sin(\varphi_{lmh})}\right]^2} \qquad T_0 = \left[\frac{|t_{lmh} t_{hml}|}{2\rho_{lmh} \sin(\varphi_{lmh})}\right]^2 \qquad (13)$$

This is a new equation similar to eq. (10), except for a multiplicative common factor that accounts for the decrease in $T$ due to absorption, $T_0(\theta) \leq 1$, that depends on the incidence angle. The independence of $T_0$ on the intracavity thickness $d$ means that eq. (11) is valid for the location of transmittance maxima in $d$ for each fixed incidence angle. Besides, it is still a very good approximation for the position of transmittance maxima in $\theta$ for a constat cavity thickness. Fig. 4 shows two examples of the good agreement of the prediction of eq. (11) for the position of the resonances and the results of a computation of transmittance using non ideal metals.

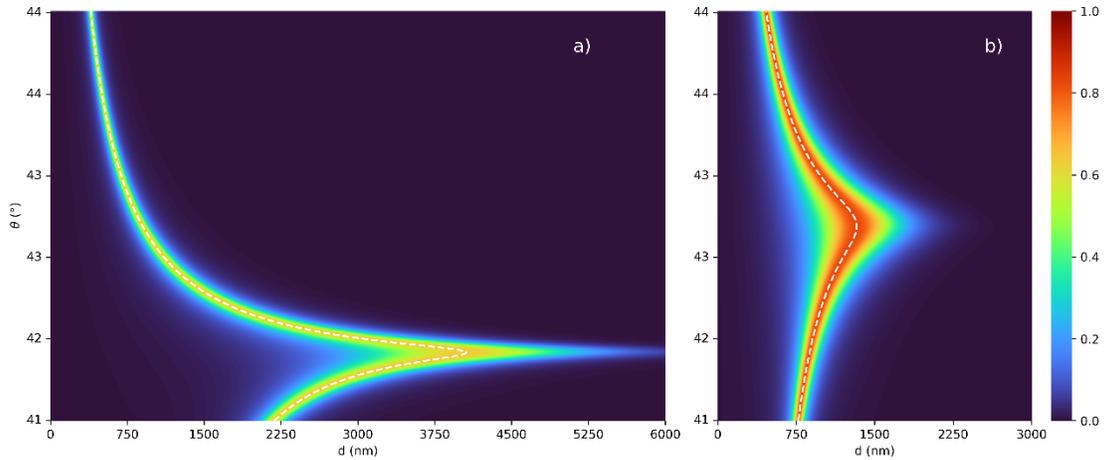

Fig. 4 Transmittance in the $(d, \theta)$ plane with $\omega_p = 1.5 \cdot 10^{16}\ s^{-1}$ and $\Gamma = 3 \cdot 10^{13} s^{-1}$ a) $\lambda = 1\ \mu m$ b) $\lambda = 0.6\ \mu m$ The white dashed lines correspond to the resonance predicted by eq. (11).

Thus, for pairs of values $(d, \theta)$ that verify eq. (11) the maximum value of transmittance will be $T_{max} = T_0(\theta) \leq 1$. Fig.3b shows the decrease in the fraction of transmitted light at resonances with respect to the ideal case. High transmittance maxima at the resonances happen when $\rho_{lmh}$ is kept small (but larger than one), which occurs for thin metallic layers or for low refractive index contrasts (between the real part of $n_m$ and $n_l$). Moreover, it should be noted from eq. (11) that smaller $\rho_{lmh}$ values mean thinner coalescence thicknesses. As we did for lossless metals, we can divide the transmission curve at resonance in three zones: i) $d \leq d_{co}$, $T_{max}(d, \theta) = T_0(\theta) < 1$, optimal but different transmission for the two CSP resonances, and higher for the fundamental one; ii) $d \gtrsim d_{co}, T_{max} = T(d, \theta_{co}) < T_0(\theta_{co})$, attenuated optical tunnelling (AOT) with moderate transmission; iii) $d - d_{co} \gg 1/k''_{l\perp}, T_{max} \cong T_0(\theta_{co}) \sin^2(\varphi_{lmh}) e^{-2k''_{l\perp}(d-d_{co})} \ll 1$ AOT with low exponentially decreasing transmission. The limit is ATR. In practice, this limit is reached when $d$ exceeds $d_{co}$ in some unities of $1/k''_{l\perp}$. We remember that the existence of region i) means that $\rho_{lmh} > 1$.

As mentioned above, the coefficient $r_{lmh}$ plays a relevant role in the appearance of resonances. Fig. 5 shows the angular dependence of the reflection coefficient $r_{lmh}$ when varying different parameters of the metallic mirrors. From eq. (4):

$$r_{lmh} = \frac{-r_{ml} + r_{mh} e^{i2k_{m\perp}s}}{1 - r_{ml} r_{mh} e^{ik_{m\perp}s}} \qquad (14)$$

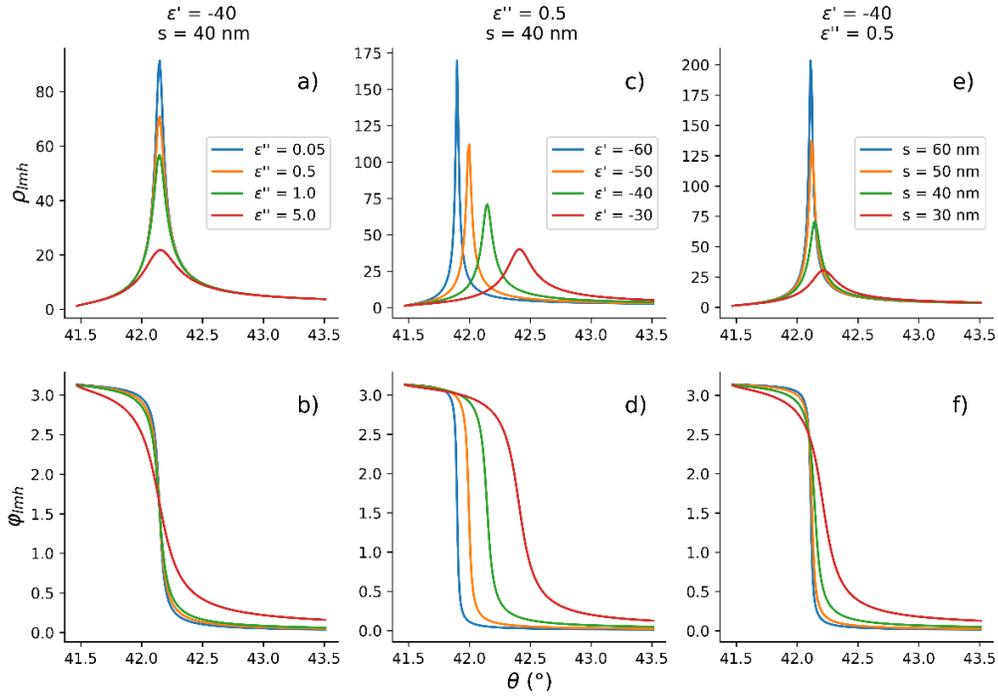

Fig. 5 Modulus $\rho_{lmh}$ (top) and phase $\varphi_{lmh}$ (bottom) of the reflection coefficient $r_{lmh}$ varying different parameters of the metallic layers $\varepsilon''$ (a, b), $\varepsilon'$ (c, d) and thickness s (e, f). $\lambda$ is set to 1000 nm.

Its modulus, shown in the subplots on top in Fig.5, rises around the coalescence angle $\theta_{co}$, reaching values far greater than unity. The peak in $\rho_{lmh}$ coincides with a phase jump, the slope of $\varphi_{lmh}$ can be seen to be maximum at that same angle in the corresponding subfigures. In Fig. 5a and 5b, increasing the imaginary part of the permittivity reduces the maximum values of the modulus and the slope of the phase, but with little change in the position of the curves. In Fig. 5c and d the effect of varying the real part of the dielectric constant is shown. An increase in the absolute value of $\varepsilon'$ also leads to an increment of both in the maximum value of $\rho_{lmh}$ and the maximum slope of $\varphi_{lmh}$, as well as to a shift of both curves towards lower angles. Finally, in Fig. 5e and f, the behaviour with increasing metal mirror thickness is seen to be like the previous case, but more moderate. It can be pointed out that as $s \to \infty$, $r_{lmh} \to r_{lm}$, which corresponds to the highest possible value of $\rho_{lmh}$ and the greatest slope of $\varphi_{lmh}$.

These changes in coefficient $r_{lmh}$ translate into changes in transmission at resonance. The decrease in the maximum values of transmittance ($T_0$) can be seen to be more drastic for higher values of the imaginary part of the permittivity (see Fig. 7b). One should also note that the coalescence thickness and angle shift slightly towards lower values with that increase of the imaginary part (see Fig,7a). We can consider that the curves obtained for $\varepsilon''/\varepsilon' = 10^{-3}$ are very close to those for a lossless metal.

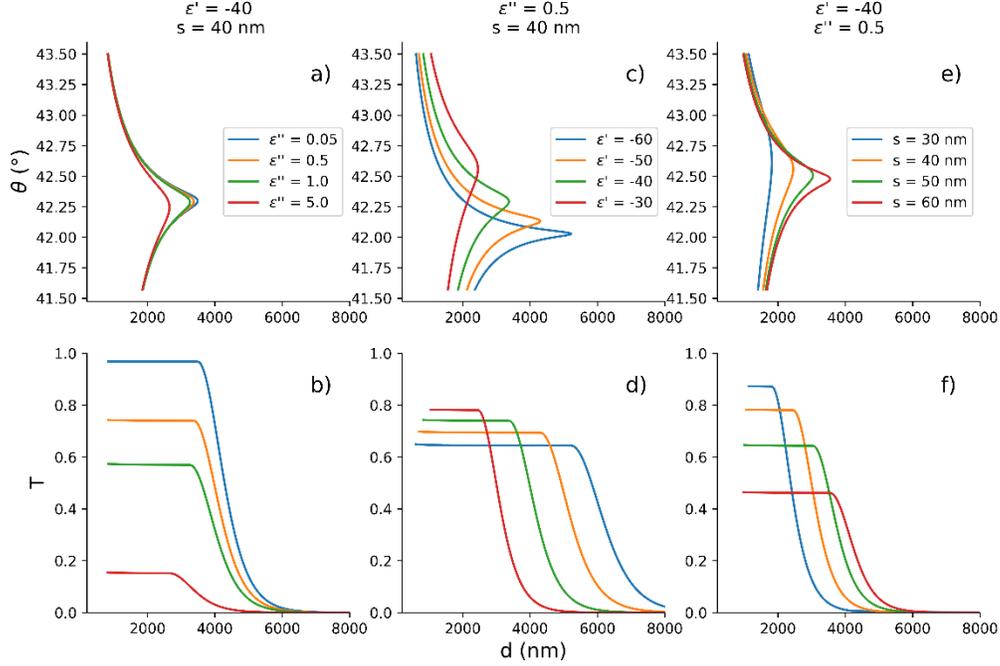

Fig. 6 Resonance curve (top) and transmittance at resonance (bottom) varying different parameters of the metallic layers $\varepsilon''$ (a, b), $\varepsilon'$ (c, d) and thickness s (e, f).

In contrast, varying the real part of the permittivity while its imaginary part is kept constant leads to a different behaviour. Maximum transmittance increases for lower absolute values of the real permittivity, since $\rho_{lmh}$ decreases. Changes in coalescence thickness and angle are more important than those in transmittance (see Fig. 6 c and d), especially if compared to the previous case when variations happened in the imaginary part. Regarding changes in transmittance as a function of the mirror thickness, $s$, (Fig. 6 e and f), $d_{co}$ increases with increasing $s$ ($\rho_{lmh}$ increases), while $\theta_{co}$ decreases. At the same, $T_0$ decreases. Therefore, for an optimal microcavity we need to maintain a balance between higher coalescence thickness and maximum transmittance.

## 5. Spectral resonance curves

So far, our analysis has considered different incidence angles but a constant wavelength. We shall now examine the opposite case, so that we can study the spectral variation of resonances. In this case, the resulting $(d, \lambda)$ map is drastically different depending on the angular region where the fixed angle, $\theta = \theta_0$, lies. Obviously, for plasmonic resonances to be observed at any wavelength $\lambda$, $\theta_0 > \theta_{cr}(\lambda)$ must be verified. Furthermore, we have a $TM_1$ resonance in the plasmonic regime for wavelengths where $\theta_{cr}(\lambda) < \theta_0 < \theta_{co}(\lambda)$, in which $\theta_{co}(\lambda)$ is the curve that defines the coalescence angle as a function of wavelength. Finally, we have a $TM_0$ resonance when $\theta_0 > \theta_{co}(\lambda)$. In Fig. 7a we show the regions where the different resonances are observed in a $(\lambda, \theta)$ map, as well as the resonance curves corresponding to two different fixed values of $\theta_0$ in a $(\lambda, d)$ map in Fig. 7b and 7c.

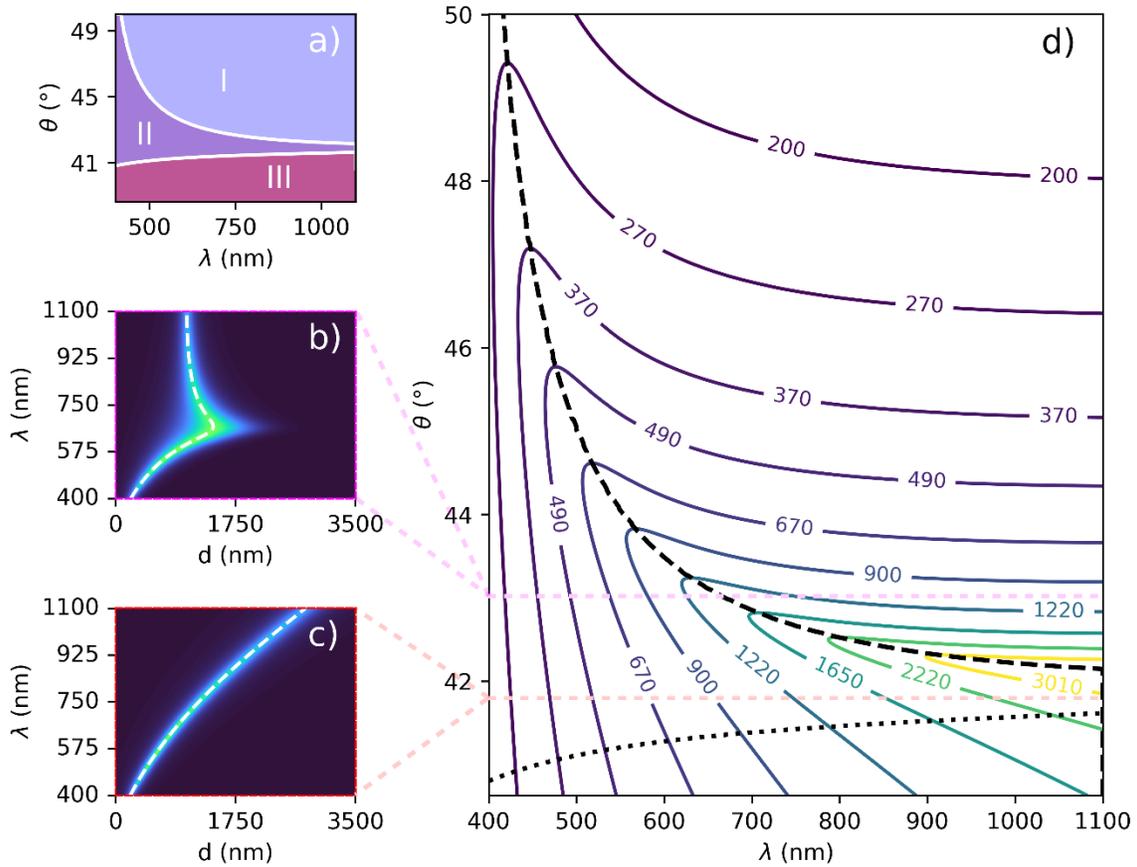

Fig. 7 (a) Diagram showing the locus of resonances. Region I corresponds to $TM_0$ resonance; region II to $TM_1$ resonances in the plasmonic regime and region III to photonic resonances ($TM_n$, n > 1). The upper continuous white line corresponds to the spectral coalescence curve $\theta_{co} = f(\lambda)$, and the lower one represents the spectral critical angle curve $\theta_{cr} = g(\lambda)$. Spectral transmittance for two different incident angles, $\theta_o = 43°$ (b) and $\theta_o = 41.76°$ (c). The dashed white line corresponds to the resonant condition (15). (d) Map of resonance using the lossy Drude model, eq. (12), with $\omega_p$ = 1.25 $10^{16}$ s$^{-1}$, $\Gamma$ = 8.1 $10^{13}$ s$^{-1}$. The metal thickness is s = 45 nm.

In any case, the plasmonic resonances must verify the resonant condition in eq. (11), equivalent to that seen in the angular case:

$$d = \frac{1}{k_{ln}''} \ln \rho_{lmh} \equiv H(\lambda, \theta_0) = G(\lambda) \qquad (15)$$

Thus, depending on the chosen fixed incidence angle, this condition might have two, one or no solutions for different wavelengths in the considered spectral range. On the other hand, let us consider the resonance curves $d = H(\lambda, \theta)$. The curves plotted in Fig. 7d were obtained using the lossy version of the Drude model and form a contour plot that allows us to track the position of the resonance along the three variables. The maxima on the plasmonic branches in Fig. 2 correspond to one of these curves. For a given thickness, corresponding to any of the curves shown in the figure, the fundamental CSP resonance is located at higher angles and wavelengths and converges with the TM$_1$ resonance at the coalescence curve, $\theta_{co} = f(\lambda)$ (the black dotted line), as can be deduced from Fig. 7a. It can be observed that the thicker the cavity, the longer the wavelength and the smaller the angle for which the coalescence curve is reached. Furthermore, the ratio between $d_{co}$ and wavelength increases with the latter. For example, $d_{co} \sim \lambda$ for a wavelength around 500 nm, $d_{co} \sim 2\lambda$ for a wavelength around 700 nm and $d_{co} \sim 3\lambda$ for a wavelength around 900 nm.

The information in figure 7d can be analysed through several interesting cuts. Thus, vertical lines correspond to slices of the graph for a constant wavelength and are associated to transmittance maxima as those shown in Fig. 4. Otherwise, horizontal slices correspond to spectral variations for a fixed angle of incidence and are associated to transmittance maxima as those shown in Figs. 7b and 7c.

## 6. Experimental

To validate the results presented in the previous sections, we conducted experimental measurements under selected configurations and compared them with the theoretical prediction. The experimental setup is illustrated in Fig 8, which shows the two configurations used: (a) for a constant wavelength analysis at different angles and thicknesses; (b) for spectral/angular analysis at a fixed intracavity thickness and. To achieve angles of incidence higher than the critical one we use prism coupling and a similar decoupling prism to measure the system transmission. The prisms are made of BK7 glass and have a thin silver layer on their hypotenuses deposited with a PVD chamber by Balzers. Layer thicknesses were measured with a needle profilometer (model Dektak 3 from Veeco Metrology), yielding results consistent with the nominal values given by the deposition system.

The light source is an NKT Supercontinuum source. Spectral measurements were performed with a custom-built grating spectrometer with a spectral range from 550 to 1050 nm and a resolution of 0.5 nm/pixel. A photodiode sensor (Thorlabs S130C) and different spectral filters were used to collect the transmitted light signal at constant wavelengths. Another photodiode controls the incident power to have a reference. After selection of TM polarization with a polarizing beam splitter, we slightly focus the light on the air gap between the prisms. For angular positioning and control, we dispose a rotation platform (model URS75BCC from Newport Optics). The two prisms were aligned in a 3D-printed capsule and brought closer together with a Thorlabs PIA25 piezoelectric actuator, which allows precise displacement steps of approximately 20 nm/step.

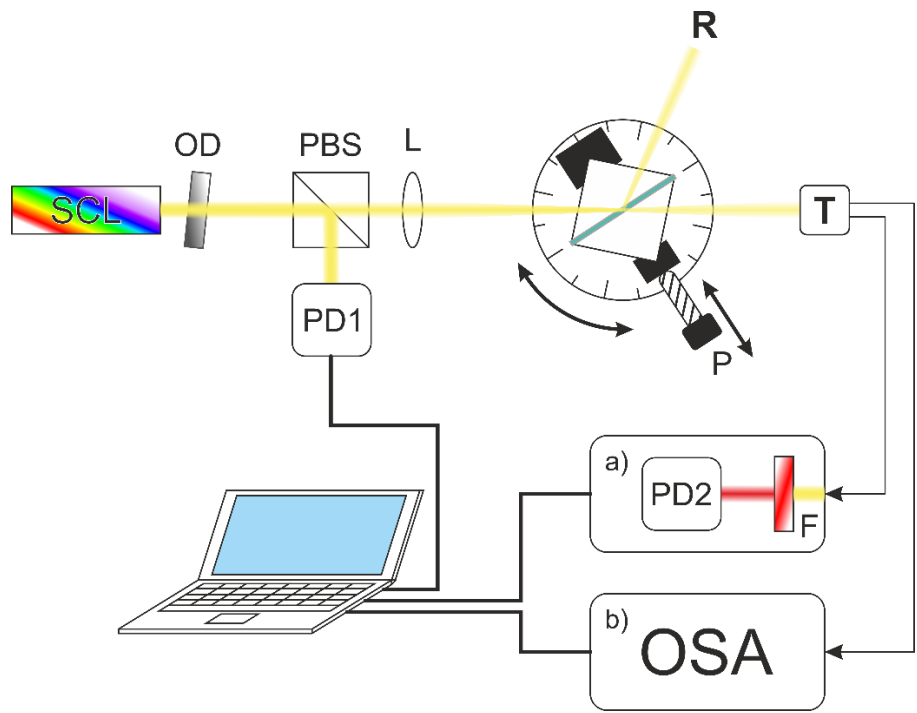

Figure 8. Schematic of the experimental setup used for θ-d (a) and spectral/angular (b) analysis. SCL supercontinuum source, OD optical density, PBS polarizing beam splitter, L converging lens, P piezoelectric actuator, F spectral filter; PD photodiode.

A comparison between simulated and experimental transmittance with an air gap of constant thickness was presented in Figure 2. The main difference between theory and experiment is a reduction in peak prominence and a broadening of the resonance peaks in the experimental data, particularly as the peaks shift away from the coalescence point. A similar trend can be observed for other cavity thickness.

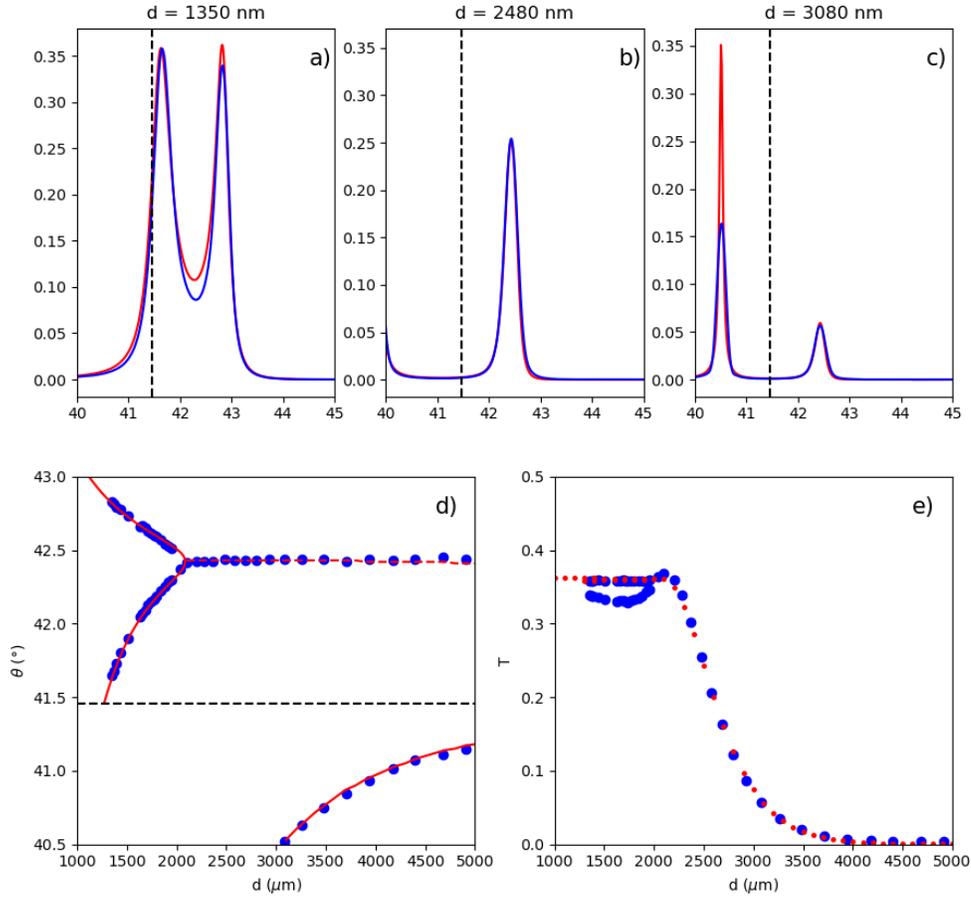

Fig. 9, Experimental measurement taken at $\lambda = 800\ nm$ and $s \simeq 39nm$. (a)-(c) Transmittance versus incidence angle at three different cavity thickness. (d) resonant angles for different thickness and their peak transmittance (e). Blue lines and dots are experimental values, while red ones correspond to theoretical predictions. The dashed black line corresponds to the critical angle.

Figs. 9(a-c) show transmittance measurements over an angular range from 40° to 45° at a fixed wavelength $\lambda = 800$ nm for a similar BK7-silver-air structure, and for three different intracavity thicknesses. The thicknesses in (b) and (c) are greater than the coalescence distance $d_{co}$, while the one in (a) is smaller. In (a) the two CSP resonances are resolved and exhibit optimal transmission; in (b) there is only a degenerate CSP resonance with moderate transmission; in (c), a photonic resonance is observed together with the degenerate CSP resonance with low transmission. As in Fig. (2) the experimental and simulated transmission differ more below the critical angle

Figs. 9(d-e) display the evolution of the resonances angular position and their transmittances as the cavity is gradually closed using the piezo actuator, until a minimum achievable spacing around 1300 nm. The curves in Fig. 9d represent the resonance peak positions in a $(d, \theta)$ map for plasmonic and first full photonic (TM$_2$) resonances. Fig. 9e shows the transmittance values for the plasmonic resonances.

These results demonstrate good agreement between experimental and simulated data. The traceability of individual peaks allows for a clear comparison. In Fig. 9(d) the coalescence of the two CSP is clearly visible, along with the emergence of the first full photonic resonance. It is necessary to point out that the permittivity of noble metals varies with film thickness at the nanometre scale [38]. Thus, the values used in our simulations for silver were fitted to match the experimental data, rather than taken directly from literature. Nonetheless, the fitted values were consistent with previously reported values [39, 40] for the taken wavelength (see table 2).

Table 2. Value of real and imaginary part of the dielectric constant of silver

| Silver dielectric constant at 800 nm | $\varepsilon'$ | $\varepsilon''$ |
|---|---|---|
| 20 nm thick film [39] | -27.1589 | 0.9798 |
| Bulk [40] | -28.7300 | 1.9165 |
| Our simulation | -28.6198 | 1.5085 |

## 7. Conclusions

In this work, we have conducted a detailed theoretical study of coupled surface plasmon (CSP) resonances in symmetric MLM optical microcavities surrounded by a high index medium (H). Starting from a generalized multi-beam interference model for three-layer systems, we extend it into the plasmonic regime where the inner wave is evanescent at high incidence angles, due to total internal reflection. This permits the definition of a clear resonance condition for high transmittance through the cavity, under conditions where strong reflection would be expected.

Our analysis demonstrates that CSP resonances, which arise from coherent oscillations at both metal-dielectric interfaces, give rise to resonant optical tunnelling. We derive and verify resonance conditions, including the concept of coalescence thickness and angle, which mark the point at which the two distinct CSP resonances degenerate into a single transmission peak. These phenomena were first analysed for lossless ideal metals and then extended to real lossy metals. The inclusion of realistic material parameters shows how losses limit transmission efficiency, but do not invalidate the resonance conditions at over-wavelength thicknesses. In the study the relevance of the amplitude coefficient $r_{lmh}$ is recognised.

We complemented our theoretical findings with experimental measurements using an all-custom-built setup. The results, obtained by precisely controlling cavity thickness and monitoring transmittance at varying wavelengths or incidence angles, show excellent agreement with the theoretical simulations. Key phenomena such as attenuated optical tunnelling and resonant coalescence were clearly observed as predicted.

These findings reinforce the potential of CSP-based microcavities for applications in highly selective optical filtering, sensing, and nanoscale light manipulation. Moreover, the simple analytical condition we provide for resonance positioning could be a valuable tool for the design of practical plasmonic devices operating in the visible and near-infrared spectral regions.

**Notes**

The authors declare no competing financial interest.

ACKNOWLEDGEMENTS

AD would like to thank Ministerio de Universidades for the financial support through FPU21/01302. YA acknowledges the funding from the postdoctoral fellowship ED481D-2024-001 from Xunta de Galicia. This research was funded by the project USC 2024-PU031 from the University of Santiago de Compostela and GRC ED431C 2024/06 from Xunta de Galicia.

REFERENCES

1. Ritchie, R. H. Plasma Losses by Fast Electrons in Thin Films. *Phys. Rev.* **106**, 874–881 (1957).

2. Turbadar, T. Complete Absorption of Light by Thin Metal Films. *Proc. Phys. Soc.* **73**, 40 (1959).

3. Welford, K. R. & and Sambles, J. R. Coupled Surface Plasmons in a Symmetric System. *Journal of Modern Optics* **35**, 1467–1483 (1988).

4. Economou, E. N. Surface Plasmons in Thin Films. *Phys. Rev.* **182**, 539–554 (1969).

5. Dragila, R., Luther-Davies, B. & Vukovic, S. High Transparency of Classically Opaque Metallic Films. *Phys. Rev. Lett.* **55**, 1117–1120 (1985).

6. Burke, J. J., Stegeman, G. I. & Tamir, T. Surface-polariton-like waves guided by thin, lossy metal films. *Phys. Rev. B* **33**, 5186–5201 (1986).

7. Homola, J. & Piliarik, M. Surface Plasmon Resonance (SPR) Sensors. in *Surface Plasmon Resonance Based Sensors* (ed. Homola, J.) 45–67 (Springer, Berlin, Heidelberg, 2006). doi:10.1007/5346_014.

8. Kretschmann, E. Die Bestimmung optischer Konstanten von Metallen durch Anregung von Oberflächenplasmaschwingungen. *Z. Physik* **241**, 313–324 (1971).

9. Otto, A. Excitation of nonradiative surface plasma waves in silver by the method of frustrated total reflection. *Z. Physik* **216**, 398–410 (1968).

10. Akimov, Y. Optical resonances in Kretschmann and Otto configurations. *Opt. Lett., OL* **43**, 1195–1198 (2018).

11. Kovacs, G. J. & Scott, G. D. Optical excitation of surface plasma waves in layered media. *Phys. Rev. B* **16**, 1297–1311 (1977).

12. Yang, F., Bradberry ,G.W. & and Sambles, J. R. Coupled Surface Plasmons at 3·391 μm. *Journal of Modern Optics* **37**, 993–1003 (1990).

13. Wang, Y. Wavelength selection with coupled surface plasmon waves. *Applied Physics Letters* **82**, 4385–4387 (2003).

14. Guo, J., Tu, Y., Yang, L., Wang, L. & Wang, B. Design of a double grating-coupled surface plasmon color filter. in *Optical Components and Materials XIII* vol. 9744 38–44 (SPIE, 2016).

15. Narayanan, V., Arora, A., Amirtharaj, S. P. & Krishnan, A. Plasmon-coupled hybrid Fabry–Pérot cavity modes in submicron metal–dielectric–metal arrays for enhanced color filtering. *OE* **58**, 057109 (2019).

16. Huang, B.-R. *et al.* Reduction of angular dip width of surface plasmon resonance sensor by coupling surface plasma waves on sensing surface and inside metal–dielectric–metal structure. *Journal of Vacuum Science & Technology A* **31**, 06F104 (2013).

17. Mohapatra, S. & Moirangthem, R. S. Theoretical study of modulated multi-layer SPR device for improved refractive index sensing. *IOP Conf. Ser.: Mater. Sci. Eng.* **310**, 012017 (2018).

18. Zhang, P. *et al.* A Waveguide-Coupled Surface Plasmon Resonance Sensor Using an Au-MgF2-Au Structure. *Plasmonics* **14**, 187–195 (2019).

19. Alwahib, A. A., Al-Rekabi, S. H. & Muttlak, W. H. Comprehensive study of generating sharp dip using numerical analysis in prism based surface plasmon resonance. *AIP Conference Proceedings* **2213**, 020143 (2020).

20. Feng, J., Okamoto, T., Simonen, J. & Kawata, S. Color-tunable electroluminescence from white organic light-emitting devices through coupled surface plasmons. *Applied Physics Letters* **90**, 081106 (2007).

21. Shimizu, K. T., Pala, R. A., Fabbri, J. D., Brongersma, M. L. & Melosh, N. A. Probing Molecular Junctions Using Surface Plasmon Resonance Spectroscopy. *Nano Lett.* **6**, 2797–2803 (2006).

22. Shin, H. & Fan, S. All-Angle Negative Refraction for Surface Plasmon Waves Using a Metal-Dielectric-Metal Structure. *Phys. Rev. Lett.* **96**, 073907 (2006).

23. Yoshida, M., Tomita, S., Yanagi, H. & Hayashi, S. Resonant photon transport through metal-insulator-metal multilayers consisting of Ag and SiO2. *Phys. Rev. B* **82**, 045410 (2010).

24. Wu, P.-T., Wu, M.-C. & Wu, C.-M. A nanogap measuring method beyond optical diffraction limit. *Journal of Applied Physics* **102**, 123111 (2007).

25. Wu, P.-T., Wu, M.-C. & Wu, C.-M. Measurement of the air gap width between double-deck metal layers based on surface plasmon resonance. *Journal of Applied Physics* **107**, 083111 (2010).

26. Liu, Y. & Kim, J. Numerical investigation of finite thickness metal-insulator-metal structure for waveguide-based surface plasmon resonance biosensing. *Sensors and Actuators B: Chemical* **148**, 23–28 (2010).

27. Elements of the theory of interference and interferometers. in *Principles of Optics: Electromagnetic Theory of Propagation, Interference and Diffraction of Light* (eds. Wolf, E. & Born, M.) 286–411 (Cambridge University Press, Cambridge, 1999). doi:10.1017/CBO9781139644181.016.

28. Hooper, I. R., Preist, T. W. & Sambles, J. R. Making Tunnel Barriers (Including Metals) Transparent. *Phys. Rev. Lett.* **97**, 053902 (2006).

29. Xiang, L. *et al*. Strong enhancement of Goos–Hänchen shift through the resonant optical tunneling effect. *Opt. Express, OE* **30**, 47338–47349 (2022).

30. Davidovich, M. V. Resonant Tunneling of Photons in Layered Optical Nanostructures (Metamaterials). *Tech. Phys.* **69**, 1521–1530 (2024).

31. Tomaš, M.-S. Recursion relations for generalized Fresnel coefficients: Casimir force in a planar cavity. *Phys. Rev. A* **81**, 044104 (2010).

32. Doval, A., Rodríguez-Fernández, C. D., González-Núñez, H. & de la Fuente, R. Fresnel coefficients, coherent optical scattering, and planar waveguiding. *Phys. Scr.* **99**, 026101 (2024).

33. Monzón, J. J., Sánchez-Soto, L. L. & Bernabeu, E. Influence of coating thickness on the performance of a Fabry–Perot interferometer. *Appl. Opt., AO* **30**, 4126–4132 (1991).

34. Smith, L. H., Taylor ,Melita C., Hooper ,Ian R. & and Barnes, W. L. Field profiles of coupled surface plasmon-polaritons. *Journal of Modern Optics* **55**, 2929–2943 (2008).

35. Wang, X., Yin, C. & Cao, Z. Symmetrical Metal-Cladding Waveguide. in *Progress in Planar Optical Waveguides* (eds. Wang, X., Yin, C. & Cao, Z.) 145–162 (Springer, Berlin, Heidelberg, 2016). doi:10.1007/978-3-662-48984-0_6.

36. Filgueira-Rama, C., Doval, A., Arosa, Y. & de la Fuente, R. Beneath and beyond frustrated total reflection: A practical demonstration. *American Journal of Physics* **93**, 157–163 (2025).

37. Ordal, M. A., Bell, R. J., Alexander, R. W., Long, L. L. & Querry, M. R. Optical properties of fourteen metals in the infrared and far infrared: Al, Co, Cu, Au, Fe, Pb, Mo, Ni, Pd, Pt, Ag, Ti, V, and W. *Appl. Opt., AO* **24**, 4493–4499 (1985).

38. Gong, J., Dai, R., Wang, Z. & Zhang, Z. Thickness Dispersion of Surface Plasmon of Ag Nano-thin Films: Determination by Ellipsometry Iterated with Transmittance Method. *Sci Rep* **5**, 9279 (2015).

39. Ciesielski, A., Skowronski, L., Trzcinski, M. & Szoplik, T. Controlling the optical parameters of self-assembled silver films with wetting layers and annealing. *Applied Surface Science* **421**, 349–356 (2017).

40. Ferrera, M., Magnozzi, M., Bisio, F. & Canepa, M. Temperature-dependent permittivity of silver and implications for thermoplasmonics. *Phys. Rev. Mater.* **3**, 105201 (2019).